\documentclass[useAMS,usenatbib]{mn2e}
\usepackage{graphicx}
\title[GEMINI near-infrared spectroscopic observations of young massive stars
embedded in molecular clouds]{GEMINI near-infrared spectroscopic observations of young massive stars
embedded in molecular clouds}
\author[A. Roman-Lopes]{A. Roman-Lopes$^{1}$\thanks{roman@dfuls.cl}, Z. Abraham$^{2}$\thanks{zulema@astro.iag.usp.br}, 
R. Ortiz$^{3}$\thanks{ortiz@astro.iag.usp.br}, A. Rodriguez-Ardila$^{4}$\thanks{aardila@lna.br}\\
$^{1}$Physics Department - Universidad de La Serena - Cisternas, 1200 - La Serena - Chile \\
$^{2}$Departamento de Astronomia, IAG/USP, Rua do Mat\~ao, 1226, Cidade
Universit\'aria, 05508-900, S\~ao Paulo, SP, Brazil\\
$^{3}$Escola de Artes, Ci\^encias e Humanidades, USP, Av. Arlindo Bettio, 1000,
03828-000, S\~ao Paulo, SP, Brazil\\
$^{4}$Laborat\'orio Nacional de Astrof\'\i sica, Rua Estados Unidos, 154,
Itajub\'a, 37504-364, Brazil}
\begin{document}

\date{}

\pagerange{\pageref{firstpage}--\pageref{lastpage}} \pubyear{2002}

\maketitle

\label{firstpage}

\begin{abstract}
$K$-band spectra of young stellar candidates in four southern hemisphere clusters have been obtained with the near-infrared spectrograph GNIRS 
in Gemini South. The clusters are associated with IRAS sources that have colours characteristic of ultracompact HII regions. Spectral types 
were obtained by comparison of the observed spectra with those of a NIR 
library; the results include the spectral classification of nine massive stars and seven objects confirmed as background
late-type stars. Two of the studied sources have $K$-band spectra compatible with those characteristic of very hot stars, as inferred from the presence of
C{\sc iv}, N{\sc iii}, and N{\sc v} emission lines at 2.078 $\mu$m , 2.116 $\mu$m, and 2.100 $\mu$m respectively. One of them, I16177\_IRS1,
has a $K$-band spectrum similar to that of Cyg OB2 7, an O3If* supergiant star.
The nebular K-band spectrum of the associated UC H{\sc ii} region shows the s-process [Kr{\sc iii}] and [Se{\sc iv}] 
high excitation emission lines, previously identified only in planetary nebula.
One young stellar object (YSO) was found in each cluster, associated with either the main IRAS source or a nearby resolved 
MSX component, confirming the results obtained from previous NIR photometric surveys. 
The distances to the stars were derived from their spectral types and previously determined $JHK$ magnitudes; they 
agree well with the values obtained from the kinematic method, except in the case of IRAS15408-5356, for which the spectroscopic distance is 
about a factor two smaller than the kinematic value.

\end{abstract}


\section{Introduction}

\begin{figure*}
\centering
\includegraphics[bb=53 190 545 645,width=16cm,clip]{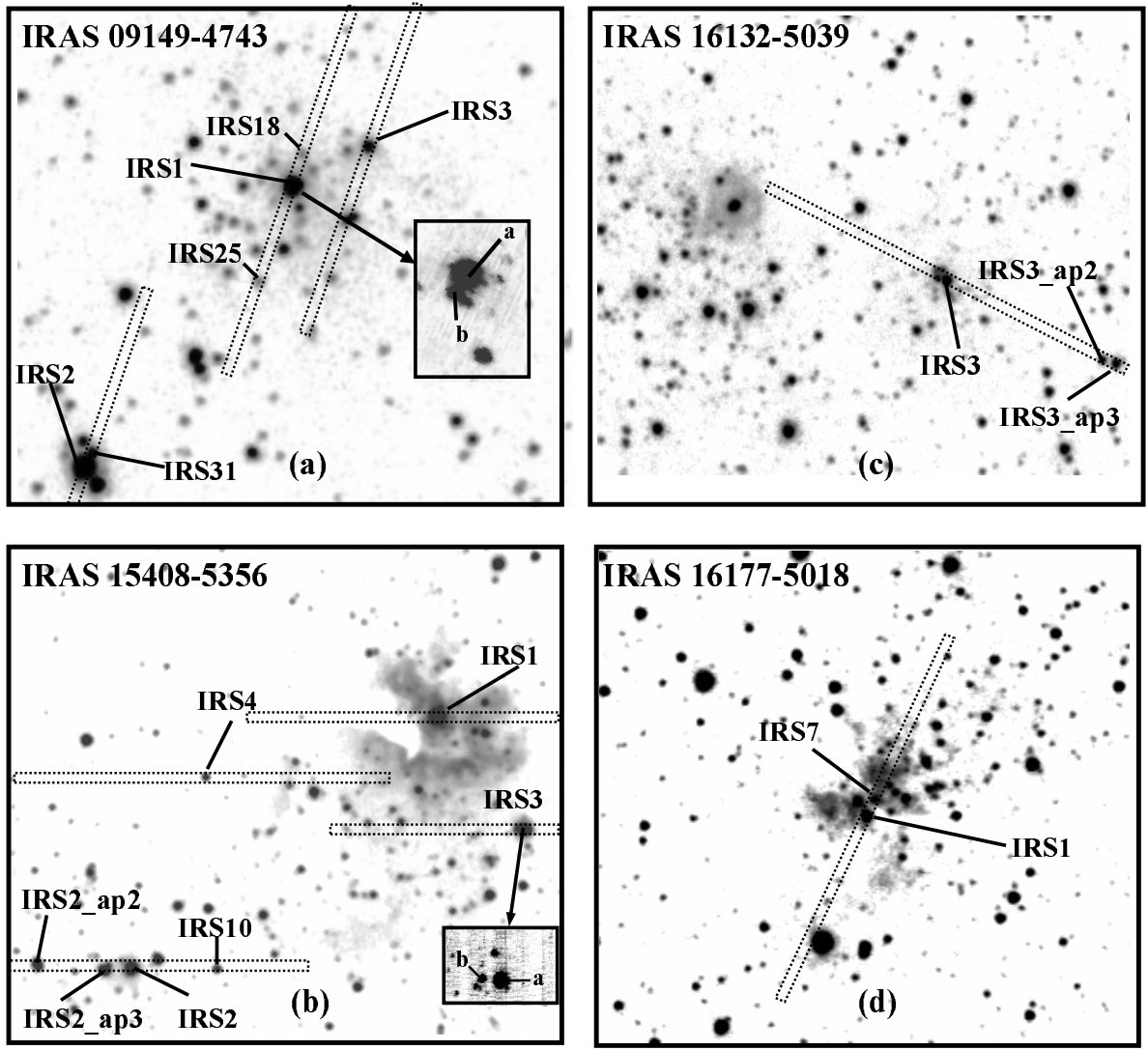}
\caption{NIR finding charts (taken from previous NIR imaging surveys) of the clusters associated to 
IRAS09149-4743 (a), 
IRAS15408-5356 (b), IRAS16132-5039 (c) and 
IRAS16177-5018 (d) sources. Each image is about 2$\arcmin$ $\times$ 2.5$\arcmin$ in size.
North is to the top, east to the left.
In each cluster, we indicate the slit position (dotted lines) and the sources for which we obtained $K$ band GNIRS spectra. 
Based on the GNIRS acquisition images, we found that IRAS09149-4743IRS1 and
IRAS15408-5356IRS3 are in fact double, as shown in the insets in panels (a) and (b), respectively, at plate scale of 0.15$\arcsec$ per pixel.
}
\end{figure*}

Massive star forming regions are commonly found embedded in high-density
molecular clouds. They can be traced, for example, by the presence of CO (Churchwell et al. 1992, 
May et al. 1993, Hofner et al. 2000), CS (Churchwell, Walmsley \& Wood 1992, Bronfman, Nyman \& May 1996), and 
NH$_3$ (Cesaroni, Walmsley \& Churchwell 1992, Cesaroni et al. 1994), transitions in the 
radio spectra of the clouds, and are often associated with methanol and water masers.  
The presence of young massive
stars still embedded in their parental molecular cloud can also be inferred from radio continuum and line surveys 
of IRAS sources selected 
according to specific colour criteria (Wood \& Churchwell 1989; Caswell \& Haynes 1987). 

Due to the large column density of gas and dust, young clusters 
are often affected by high visual
extinction, sometimes over 30 magnitudes, which make them undetectable  at optical
wavelengths.
Dutra et al. (2003) identified many previously unknown stellar cluster candidates in the Near-Infrared 
(where the extinction
is about a tenth of that in the visual window), using the
Two Micron All Sky Survey (2MASS). The 2MASS images have been used not only to discover new young clusters, but also
to obtain physical properties of their members (Borissova et al. 2003, Leistra et al. 2005), although the limited
spatial resolution ($\sim 2"$) restricts its usage to non-crowded
regions. Several authors have taken advantage of the higher spatial resolution
provided by the new generation of near-infrared (NIR) array detectors to 
study  clusters with high stellar surface density (Horner, Lada \& Lada 1997, Gomes \& Kenyon 2001, 
Hanson, Luhman \& Rieke 2002, Massi, Lorenzetti 
\& Giannini 2003, Balog et al. 2004, Kumar, Kamath \& Davis 2004, Lada \& Muench 2004, Whitney et al. 2004, 
Figuer\^edo et al. 2002, 2005, Arias, 
Barba \& Morrell 2007, Barba \& Arias 2007).

Roman-Lopes, Abraham \& L\'epine (2003); Roman-Lopes \& Abraham (2004a,b),
Roman-Lopes \& Abraham (2006a,b), Ortiz, Roman-Lopes \& Abraham (2007), Roman-Lopes (2007), used the NIR camera CamIV, attached to the telescopes 
of the Pico
dos Dias Observatory, to study clusters associated with high density molecular clouds, and IRAS sources with colours of ultracompact HII regions. 
In these studies, \emph{JHK} colour-colour (C-C) and
colour-magnitude (C-M) diagrams were used to select the cluster member candidates. Their spectral types were 
estimated by dereddening 
magnitudes and colours in the C-M diagram up to the point where they intercept the ZAMS line. This 
method depends on an accurate knowledge of at least three parameters: (1) the distance to the cluster; (2) 
the extinction law
in that direction, and (3) the excess emission due to the possible presence of circumstellar material 
around each individual star, 
which affects mainly the $K$-band magnitudes.

\begin{figure*}
\vspace{80pt}
\centering
\includegraphics[bb=44 540 354 748,width=14cm,clip]{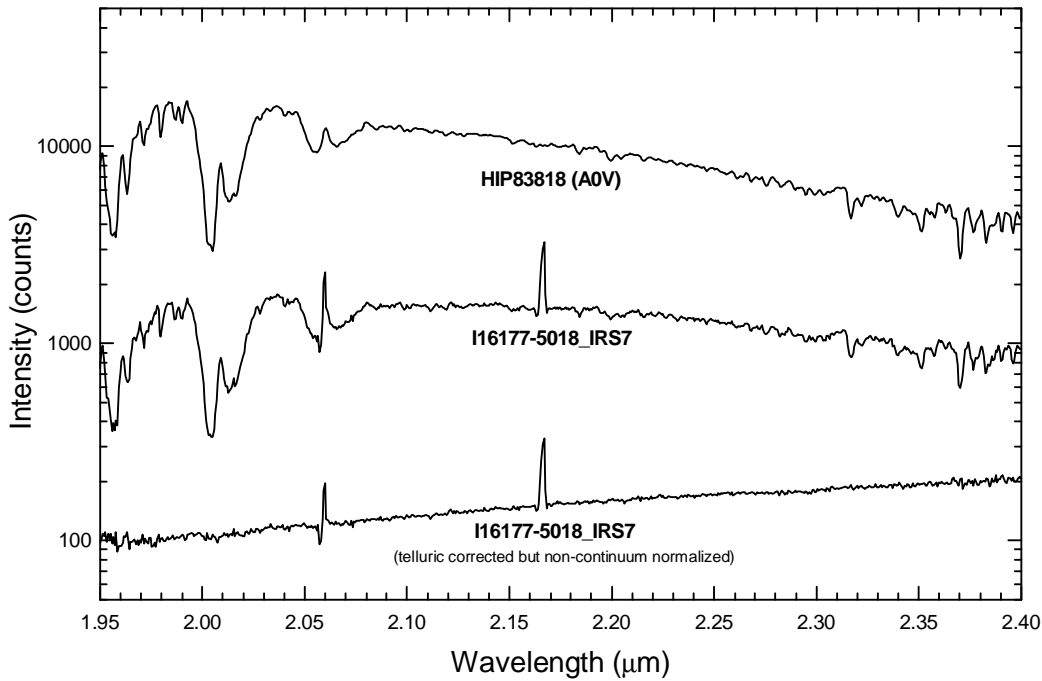}
\caption{Spectra of the A0V standard star HIP83818, the science target IRAS16177-5018\_IRS7, and
the resulting corrected spectrum (non-continuum normalized)}
\end{figure*}

Most studies of embedded clusters assume kinematic or statistical
distances and make use of a standard  interstellar extinction law (Rieke \& Lebofsky
1985), even though deviations have been reported in dense molecular
clouds (Tapia 1981, Indebetouw et al. 2005, Nishyama et al. 2006). In addition to that,  it is well established that massive young stellar  
objects (YSOs) have large infrared excess due to the presence of warm circumstellar dust (Grasdalen et al. 1975, Lada \& Adams 1992),
which will be reflected in an incorrect estimation  of their
spectral types. To circumvent part of these problems, spectral types of some of the cluster members  can be determined from $K$-band 
spectroscopy and used to determine the  distances to the clusters assuming different extinction laws.

In this work, we present $K$-band spectra of a sample of southern massive star candidates obtained with the 
Gemini Near Infrared Spectrograph (GNIRS). 
All targets were selected from previous NIR studies (Roman-Lopes et al. 2003, Roman-Lopes \& Abraham 2004a,b, 
Ortiz et al. 2007). 
With the present study we intend to obtain spectral classification of the most massive candidates in these 
clusters, and from them 
the cluster distances using the spectroscopic parallax technique. These distances are compared to 
those derived from kinematical 
methods, and can be  eventually  used to improve the rotation model of our Galaxy.

The paper is organized as follows: Section 2 presents a summary
of the four studied regions; Section 3 reports the observations,
data reduction and results. In Section 4 we discuss the results obtained for each cluster and in Section 5 
we present our conclusions.

\section{The stellar clusters}

The sources chosen for this work belong to  young stellar clusters in the southern hemisphere, which are 
associated with the 
IRAS09149-4743, IRAS15408-5356, IRAS16132-5039 and IRAS16177-5018 sources, previously  studied using infrared 
imaging techniques. 

The first cluster, associated with IRAS09149-4743, belongs to the Vela Molecular
Ridge (VMR), and is probably related to the optical HII region RCW 41. Ortiz et al. (2007) obtained \emph{JHK}
photometry of this cluster  at 1.3'' spatial resolution. The authors suggested  two stars as the more
likely candidates to ionise the nebula:  IRS1,
located at the centre of the IRAS error ellipse, and  IRS2, a member of a small ``sub-cluster'' 
containing 6 stars, situated 1.1'
southeast of the IRAS position and associated to the MSX6C-G270.2795+00.8353 source. They also 
found another bright source, IRS3, 
the reddest object in the cluster. 
The $K$-band image of the region obtained by Ortiz et al. (2007) is presented in Figure 1a, where 
these three program stars are 
identified, together with other stars that fell into the spectrograph slit.

\begin{figure*}
\vspace{10pt}
\includegraphics[bb= 54 70 504 760,width=8cm,clip]{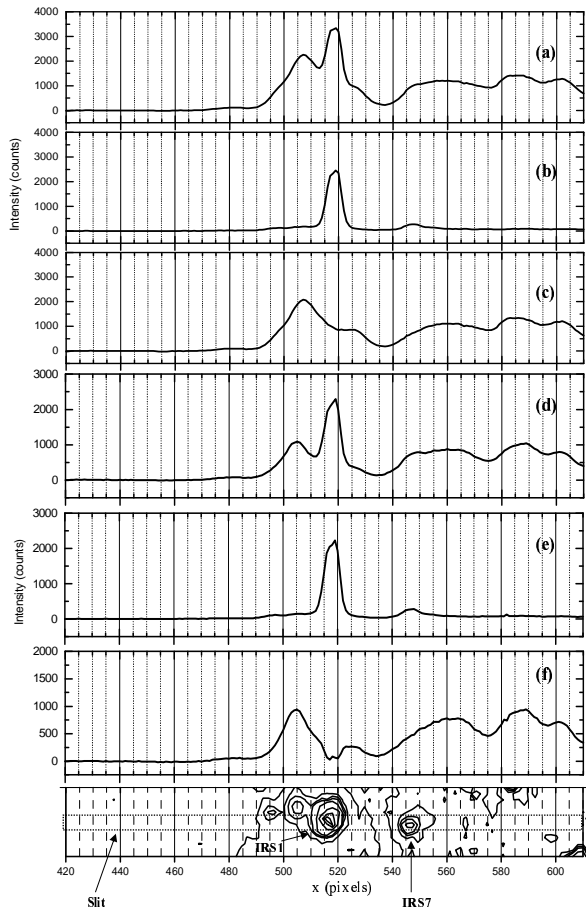}
\caption{Plots of the intensity of the Br$\gamma$ and HeI line emission as function of the spatial coordinate. 
The emission profile is shown in (a), the continuum emission profile in (b) and the resulting continuum subtracted Br$\gamma$ line is shown in (c). The analogous for the HeI line are shown in (d), (e) and (f), respectively. At the bottom we also
present the contour diagram made from the GNIRS acquisition image. There we indicate the IRS1 and IRS7 sources, and the position of the 
slit together with the scale that represents the spatial coordinate (measured in pixels).
From the analyses of this diagram we were able to estimate the amount of nebular contamination and to correct the 
stellar spectrum from it.}
\end{figure*}

The second cluster is associated with the IRAS
source 15408-5356 and the HII region RCW 95; it is seen against the
Sagittarius-Carina and Scutum-Crux spiral arms.
This cluster was studied in detail by Roman-Lopes \& Abraham (2004b), who
obtained \emph{JHK} photometry of the sources in the region. A nebula is clearly visible in
their images, with a clump of embedded stars that
include two of the probable ionising sources of RCW 95, IRS1 and IRS3, as shown in Figure 1b.

The third stellar cluster, is located towards the
IRAS source 16132-5039 and is associated
with the HII region RCW 106. The infrared image obtained by Roman-Lopes \& Abraham (2004a) shown in Figure 1c,
reveals a spheroidal nebula containing a bright star at its centre, IRS1, and a smaller concentration of stars 
to the south-west of the IRAS source, 
which coincides with the mid-infrared source MSX5C G332.5302-00.1171;
the brightest  source in this sub-cluster is labeled IRS3.

The fourth cluster is associated with IRAS 16177-5018 and, together with
IRAS 16132-5039, is embedded in the H{\sc ii} region 
RCW 106. Roman-Lopes et al. (2003), obtained \emph{JHK} images and photometry
of the  reddest stars, which have  visual extinction exceeding 28 magnitudes.
The brightest source is IRS1, located at the centre of the infrared nebula, which together with IRS7, 
seem to play a key role in the energy balance 
of the compact H{\sc ii} region (Roman-Lopes et al. 2003). An image of the region, with the two stars  
indicated, is shown in Figure 1d.

\begin{figure*}
\centering
\includegraphics[bb=86 140 525 620,width=15cm,clip]{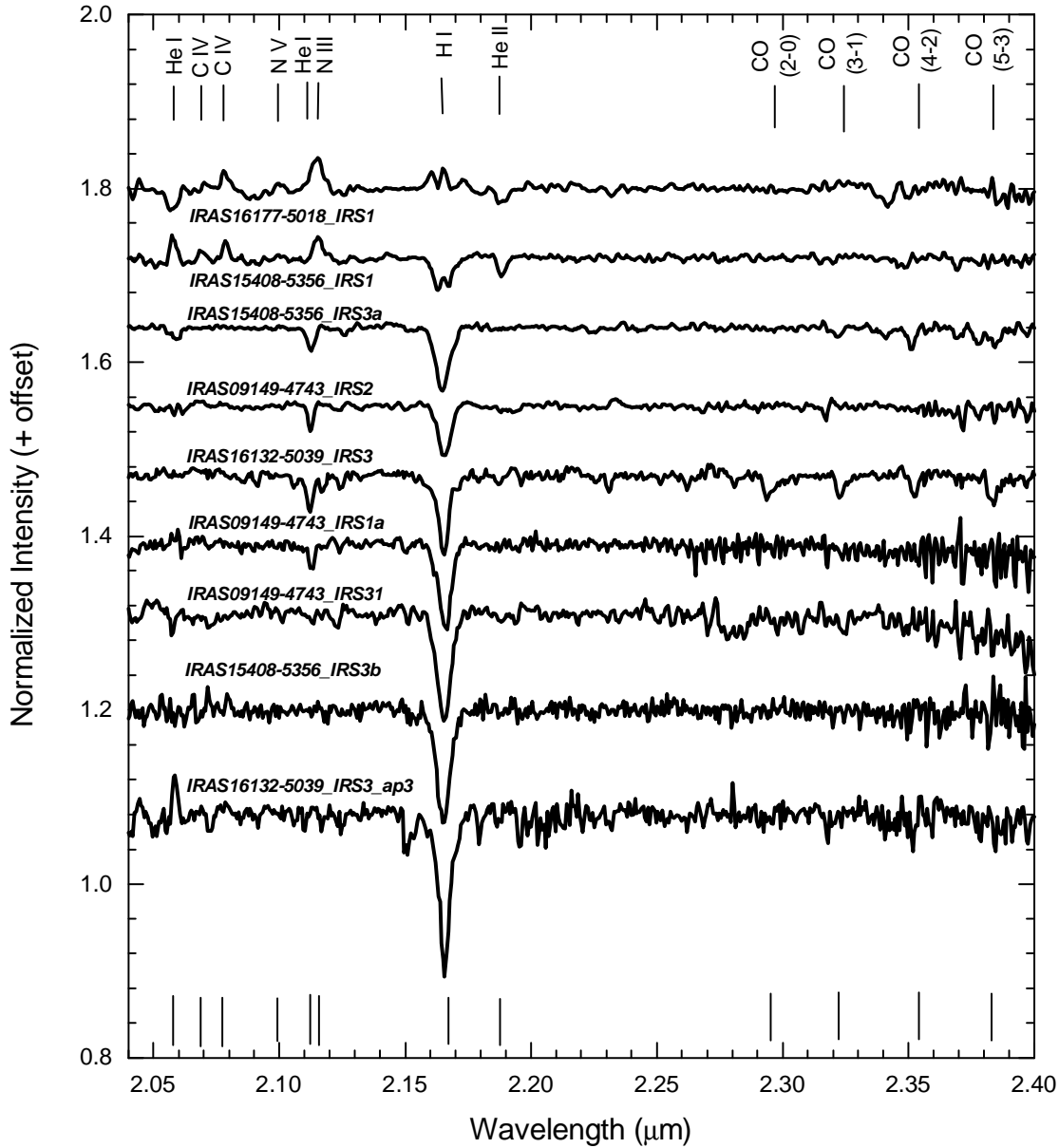}
\caption{GNIRS K-band spectra of the hot stars detected in this work. All spectra were flux normalized and are at the same scale.}
\end{figure*}

\section{Observations and results}

\subsection{GNIRS data}

$K$-band spectra of the point sources indicated in Figure 1 were
obtained on different dates using GNIRS (Elias et al. 2006) on the 8-m Gemini South
telescope at Cerro Pachon, Chile. Table 1 shows the log of observations. In all
cases the short camera with a pixel scale of 0.15$\arcsec$ pixel$^{-1}$ and a
slit size 0.675$\arcsec \times 99\arcsec$ was employed. The resolving power
of this configuration is 1600, with a theoretical wavelength coverage ranging from 1.90 to 2.50 $\mu$m.

In order to obtain good results in the sky subtraction process, the standard ABBA nodding technique 
was used to acquire the spectra. The large slit length (99\arcsec) enabled us to perform
large node shifts with the target still on the slit. This is specially useful when observing regions
where the nebular lines (like Br$\gamma$) are spatially extended, to avoid the overlap of AB positions 
that would create artifacts in the sky subtracted images.
Individual exposure times at each nod position was 3 minutes for the science targets; total exposure times are
listed in column 5 of Table 1. In order to correct the science spectra for the effect of telluric atmospheric
absorption, a nearby A0{\sc v} spectroscopic standard star was observed at similar
airmass (column 7 of Table 1), before or after the set of exposures of each science target.

The GNIRS data were reduced using the GEMINI package within IRAF\footnote{IRAF is distributed by the National Optical Astronomy Observatories, 
which are operated by the Association of Universities for Research in Astronomy, 
INC, under cooperative agreement with the National Science Foundation}.
First, the 2D {\it K}-band frames were sky-subtracted for each pair of
images taken at the two nod positions A and B, followed by division
of the resultant image by a flat-field. 
Multiple exposures for each source were combined, followed by
one-dimensional extraction of the spectra. 
Eventually, wavelength
calibration was performed using sky lines;
the typical error (1-$\sigma$) for this calibration was $\sim$5~\AA.
Telluric atmospheric
correction using the spectroscopic standard stars completed the
reduction process. 
In this last step, we divided the target spectra by the spectrum of the A0{\sc v} spectroscopic standard star,
already free of photospheric features. In the standard star, the Br$\gamma$ absorption 
is the only feature
present in the {\it K}-band spectrum. It was carefully removed by interpolation across its wings using 
continuum points on either
side of the line while its core was modeled using a Voigt profile.
In order to assure good cancellation of the telluric bands, the IRAF task
{\it telluric} was employed. The algorithm interactively
minimizes the RMS in specified wavelength regions by shifting
and scaling the target relative to the standard spectrum  to best divide
out telluric features present in the former. Shifting account
for small errors in the dispersion zero-points, while the
intensity scaling corrects for differences of airmass and
variations in the abundance of the telluric species. Typical
values of the shifts were a few tenths of a pixel (equivalent
to $\sim$2\,\AA) and the scaling factors were less than 10\%.
As an example, Figure 2 shows the spectra of the A0{\sc v} standard star HIP83818 (already free of the Br$\gamma$ line), 
the science target IRAS16177-5018\_IRS7, and its
spectrum corrected from the effects of the telluric absorption bands.

\subsection{Correcting the science spectra for nebular contamination}

As can be inferred from Figures 1b and 1d, the regions associated to IRAS15408-5356 and IRAS16177-5018 present 
strong extended emission. In fact,
the $K$-band spectra of IRAS15408-5356\_IRS1, IRAS16177-5018\_IRS1, and IRAS16177-5018\_IRS7 are contaminated by 
Br$\gamma$ and HeI (2.058 $\mu$m) 
nebular components.
Since these lines play a fundamental role in the spectral classification, they must be carefully subtracted from the
stellar spectrum. To do that,
we evaluated the nebular contribution at the position of the point sources by
studying the intensity profiles of the Br$\gamma$ and HeI (2.058 $\mu$m) lines as function of the position along the 
slit.

In Figure 3 we illustrate the procedure used in the case of the sources belonging to the IRAS16177-5018 cluster.
At the bottom panel of this figure, we present the contour diagram made from the GNIRS acquisition image; there we 
indicate the position on the slit of IRS1 and IRS7, with the spatial coordinate scale measured in pixels.
Figures 3a  and 3d show the spectral intensity at the wavelength of the Br$\gamma$ and HeI 2.058 $\mu$m 
lines respectively, which include the continuum emission. After the subtraction of the continuum, taken at 
$\pm$30 \AA (about 6 pixels) off the line centre and shown in Figures 3b and 3e, we obtained the 
Br$\gamma$ and HeI 2.058 $\mu$m line contributions, presented in Figures 3c and 3f, respectively, 
where we can see the nebular emission at both sides of the stars. In Figure 3f HeI 2.058 $\mu$m is 
clearly seen in absorption at the position of IRS1, while no absorption is present in Br$\gamma$ in 
Figure 3c. This procedure was used to obtain the spectra of all stars embedded in ionised clouds, 
even when the nebular contribution was small.

\begin{figure}
\vspace{25pt}
\centering
\includegraphics[bb=76 265 492 717,width=8.2cm,clip]{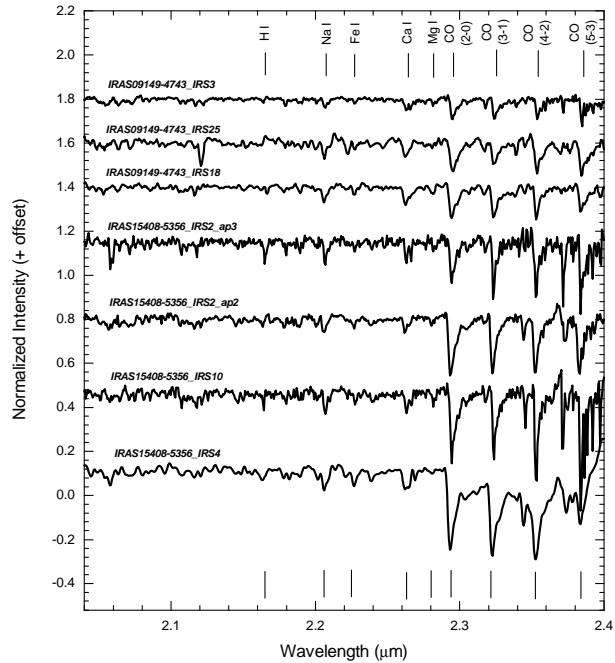}
\caption{GNIRS spectra of the late-type stars detected in this work. All spectra were continuum flux normalized and are at same scale.}
\end{figure}

\begin{figure}
\vspace{10pt}
\centering
\includegraphics[bb=37 158 335 768,width=7.5cm,clip]{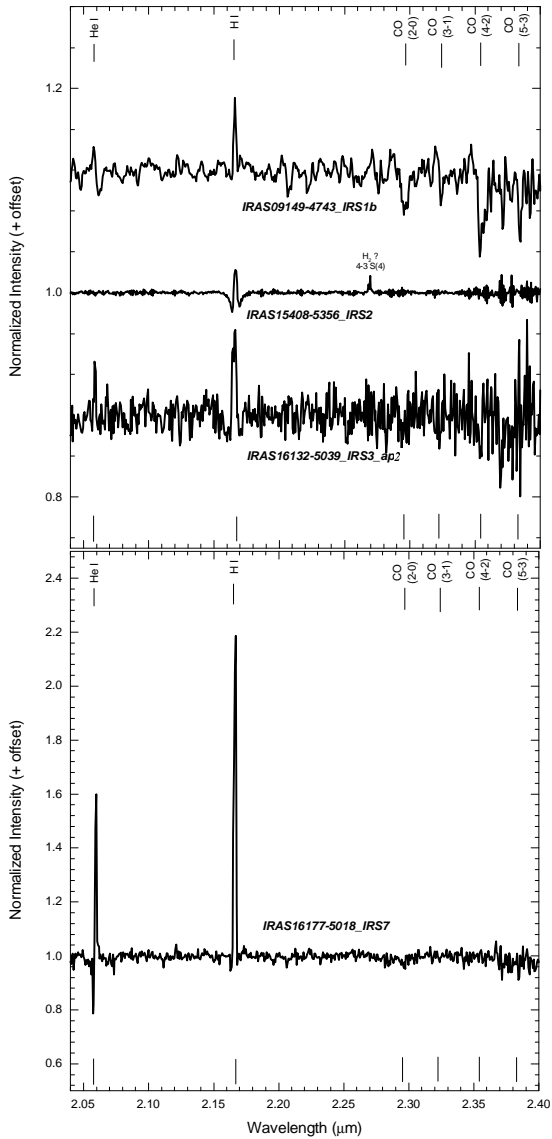}
\caption{GNIRS spectra of YSO's detected in this work. All spectra were continuum flux normalized.}
\end{figure}

\subsection{Spectral classification}

The GNIRS spectra have been organized in three groups. In the first one, the detected lines are characteristic of hot stars 
(Br$\gamma$, He{\sc i} 2.058 $\mu$m, He{\sc i} 2.113 $\mu$m, He{\sc ii} 2.185 $\mu$m, C{\sc iv} 2.078 $\mu$m, N{\sc iii} 2.116 $\mu$m, among others.); 
they are shown in Figure 4 in order of increasing Br$\gamma$ equivalent width (EW).

The spectra of the sources in the second group, shown in Figure 5, present weak Br$\gamma$, metallic (Na, Ca, Mg, etc.), and molecular (CO overtones) 
lines, characteristic of late-type stellar spectra. Finally the third group shows H and He lines in emission but no photospheric 
absorption lines, which is characteristic of young stellar objects (YSOs); their spectra are shown in Figure 6.

The library of $K$-band early-type spectra compiled by Hanson, Conti \& Rieke (1996) was used  to classify the stars in our sample. 
Since both spectra have similar resolution, the classification of the stars in group 1 was made by direct comparison. For each star, we show 
the library spectrum that best matches the target one, as well as earlier and later type library spectra (Figures 7, 8, 9 and 11).

The derived spectral types are presented in Table 2, together with the absolute 
$M_J$ magnitudes. They were computed 
considering the $M_V$ magnitudes taken from Hanson et al. (1997) and Walborn (2002), and the intrinsic 
colours given by Koornneef (1983) transformed into the 2MASS photometric system. 
Columns 6 and 7 list the intrinsic $(J-H)_0$ and $(H-K)_0$ colours corresponding to each spectral type. Columns
8, 9, 10 and 11 list the measured $J$ and $K$ magnitudes, 
as well as the $(J-H)$ and $(H-K)$ colours obtained from
previous works (Roman-Lopes et al. 2003, Roman-Lopes \& Abraham 2004a, 2004b, Ortiz et al. 2007).

\subsection{Determination of the distances}

The distances $d$ to the stars were determined using absolute and measured magnitudes presented 
in Table 2, in the equation:

\begin{equation}
m_{\lambda}-M_{\lambda} =  5\log[d{\rm (pc)}] - 5 + A({\lambda})
\end{equation}

The absorption $A({\lambda})$ is related to the colour excesses ${\rm E}(J-H)$ and ${\rm E}(H-K)$ through the 
functions $F_J(R)=A(J)/{\rm E}(J-H)$ and
$F_K(R)=A(K)/{\rm E}(H-K)$ that depend on the ratio of the total to selective extinction $R=A_V/{\rm E}(B-V)$.
Using the interstellar extinction laws given by Fitzpatrick (1999), it is possible to obtain the ratios 
$A_J/A_V = f_J(R)$, $A_K/A_V = f_K(R)$, 
${\rm E}(J-H)/{\rm E}(B-V) = g_J(R)$ and ${\rm E}(H-K)/{\rm E}(B-V) = g_K(R)$ from which the functions 
$F(R)= Rf(R)/g(R)$ can be derived.

A source of uncertainty in the distance determination originates from the use of the standard interstellar 
extinction law, which is represented by $R=3.1$, though it is common to find $R$ in the range $2.8-5.8$ along 
the galactic plane (Johnson 1965, 
Tapia 1981, Fitzpatrick 1999, 
Indebetouw et al. 2005, Nishiyama et al. 2006). 
This effect is probably produced by differences in metallicity and grain size distribution (Savage \& Mathis 1979). 
It was taken into account by using $R=2.8$, 3.1 and 5.0, and the distance to each star was calculated as the average 
of these individual values.

Another important source of error comes from the spectral type determination itself, which in our work 
has an uncertainty of about 
$\pm$ one sub-type. For O-type stars it represents about 0.2 magnitudes in the NIR, whereas for early-B 
stars it is about 0.6 magnitude 
(Hanson et al. 1997).
The only exception is the I16177-5018\_IRS1 source, for which the true luminosity class is an important source of 
uncertainty, this issue will be discussed in detail in Section 4.

A third source of uncertainty  results from the fact that young stars can be surrounded by disk and/or 
dust cocoons, which produce excess 
emission in the NIR (Grasdalen et al. 1975, Glass 1979, Lada \& Adams 1992), especially in the $K$-band. 
In order to minimize this effect, 
we used both the $J$ and $K$ band photometry to derive distances.

Table 3 shows the values of E($J-H$), E($H-K$),  $A_J$  and $A_K$ for three values of total to selective 
extinction ratios (2.8, 3.1 and 5.0).The stellar distances $d$ were computed as the average of the distances 
obtained from the $J$ and $K$ magnitudes, and the three extinction laws. The quoted errors for the  distances
are the standard deviation of the averaged values, which include the errors associated to the 
photometry, interstellar extinction law, and spectral type, as discussed above.
Finally, the distance to each cluster (Column 13) was computed as the average of the distances to each star in the cluster; the total error 
of these values is the square root of the sum in quadrature of the individual errors.

\begin{figure}
\vspace{10pt}
\centering
\includegraphics[bb=45 103 234 758,width=6.5cm, clip]{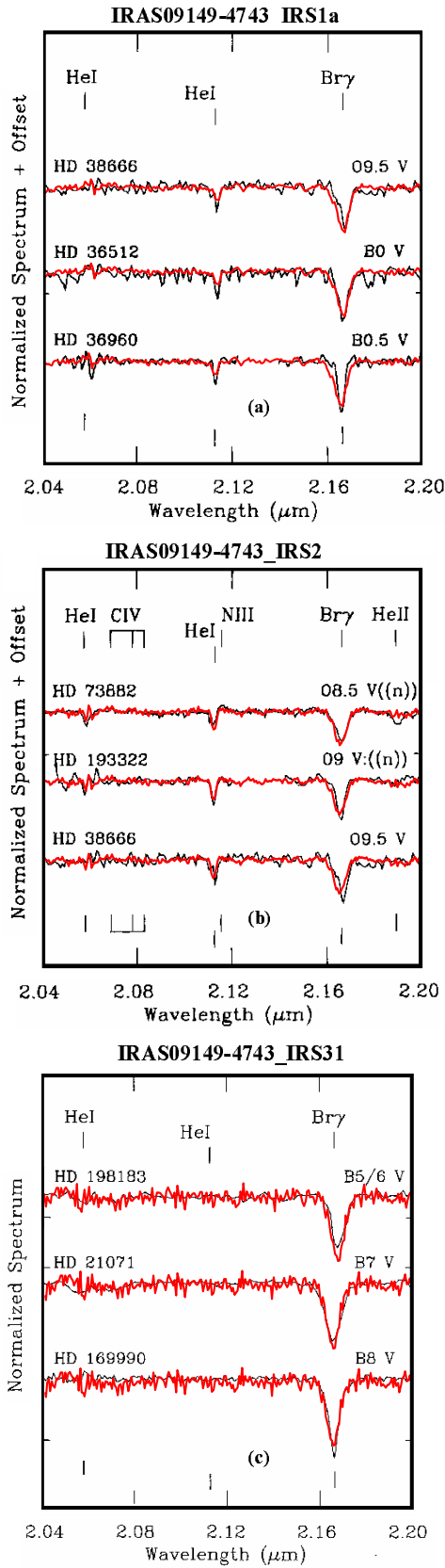}
\caption{GNIRS $K$ band spectra of IRAS09149-4743 sources (red lines) overlayed on
$K$-band spectra (black lines) taken from the library of Hanson et al. (1996).}
\end{figure}

\section{Discussion}
 
In this section, we discuss our main results and compare them 
with those obtained in the previous works.

\subsection{The IRAS09149-4743 cluster}

In this region, three program stars were chosen for observation: IRS1, IRS2 and IRS3. In addition, 
three other sources located nearby had 
their spectra taken as well: IRS18, IRS25, and IRS31 (Ortiz et al. 2007). Their relative positions are shown in Fig.1a.
The source labeled as IRS1  was resolved into two stars in this work: IRS1a and IRS1b 
(see the inset in Fig. 1a). The spectra of IRS1a and IRS2, shown in Fig. 4, exhibit two important diagnostic 
lines commonly found in  stars earlier 
than B1{\sc v}: H{\sc i} Br-$\gamma$ at 2.166 $\mu$m and the He{\sc i} line at 2.113 $\mu$m.
The source IRS1a was classified as a B0{\sc v} star (Figure 7a), affected by 9 magnitudes of visual 
extinction (Table 3), which places it at 
a distance of about $1.20\pm 0.12$ kpc. 

IRS2, belonging to the nearby sub-cluster located 1.1 arcmin to the southeast of IRS1, 
is probably an O9{\sc v} star (Figure 7b) affected by about 7$-$8 visual magnitudes; it is located at $1.27\pm 0.13$ 
kpc, virtually the same distance 
as IRS1a. IRS31 belongs to the same sub-cluster as IRS2. Its spectrum (Fig. 7c)
exhibits a strong Br-$\gamma$ line, characteristic  of late-B and early-A stars. 
This source has been classified as a B7{\sc v}-B8{\sc v} star, suffering 5-8 magnitudes of extinction in the $V$ 
band (Table 3). Its
distance is about $1.37 \pm 0.19$ kpc, reinforcing its membership. 

The distances to the three massive stars studied in this region are similar, resulting in a mean cluster distance of 
$1.3 \pm 0.2$ kpc. 
This value can be compared with the photometric distance of $0.7 \pm 0.2$ kpc quoted by Liseau et al. (1992) for 
cloud A in the VMR complex, 
and the 
kinematic distance of 1 kpc inferred from CO observations by Murphy \& May (1991). 
The good agreement between our result and that obtained from the Galactic rotation curve is notable, considering 
the complexity
of the VMR and the small radial velocity resulting from its position close to $l=270\degr$.

In Figure 6 we can see that IRS1b shows no photospheric spectral lines, but Br$\gamma$ in emission and some 
CO overtone band-heads in absorption, 
characteristic of YSO's (Casali \& Eiroa 1996, Hoffmeister et al. 2006). These CO lines are believed to be 
formed in a warm and dense 
circumstellar shell, possibly a relic of a former accretion disk. 
They can be seen in absorption or emission depending on the disc opacity, which in 
turn depends on the mass accretion rate, 
as shown by Calvet et al. (1991). Ortiz et al. (2007) also pointed out that IRS1a+b shows intense infrared 
emission beyond 5 $\mu $m, usually attributed to warm dust. 
Based on the present data, we can now state that the observed infrared excess comes from IRS1b, the dusty 
nearby companion of the IRS1a source.

IRS3 is a highly reddened IR source (Ortiz et al. 2007), located at about 1
arcmin west of IRS1 (Figure 1a). Its spectrum, shown in Figure 5, exhibits
metallic and molecular absorption lines, such as the CO (2,0) and (3,1)
transitions at 2.29 and 2.32 $\mu$m respectively, as well as the Ca{\sc
i} and Na{\sc i} lines at 2.21 and 2.26 $\mu$m, typical of K and M giant stars. 
Assuming its spectral classification as early-K, we can estimate a
lower limit for its distance. Its absolute magnitude and
intrinsic colour index would be $M_{\rm J}=-1.9$ and $(J-H)=0.62$,
which implies a colour excess ${\rm E}(J-H)=1.42$. If one assumes
$R=3.1$ then $A_{\rm J}=4.34$ and $d= 2.5$ kpc.
On the other hand, if $R=5.0$, $A_{\rm J}=4.76$, and $d=2.0$ kpc.
In any case, the distance to this star would be twice
as large as the distance to RCW 41. Besides IRS3, two other stars
in the neighbourhood have been classified as late-type:
IRS18 and IRS25 (Fig. 5). Similarly to IRS3, they occupy a position
in the $(J-H)$ versus $(H-K)$ C-C diagram consistent with their classification
as late-type stars, and their $JHK$ magnitudes given in Ortiz et
al. (2007) imply that they must also be background objects.

\begin{figure}
\vspace{10pt}
\centering
\includegraphics[bb=48 83 244 765,width=6.5cm,clip]{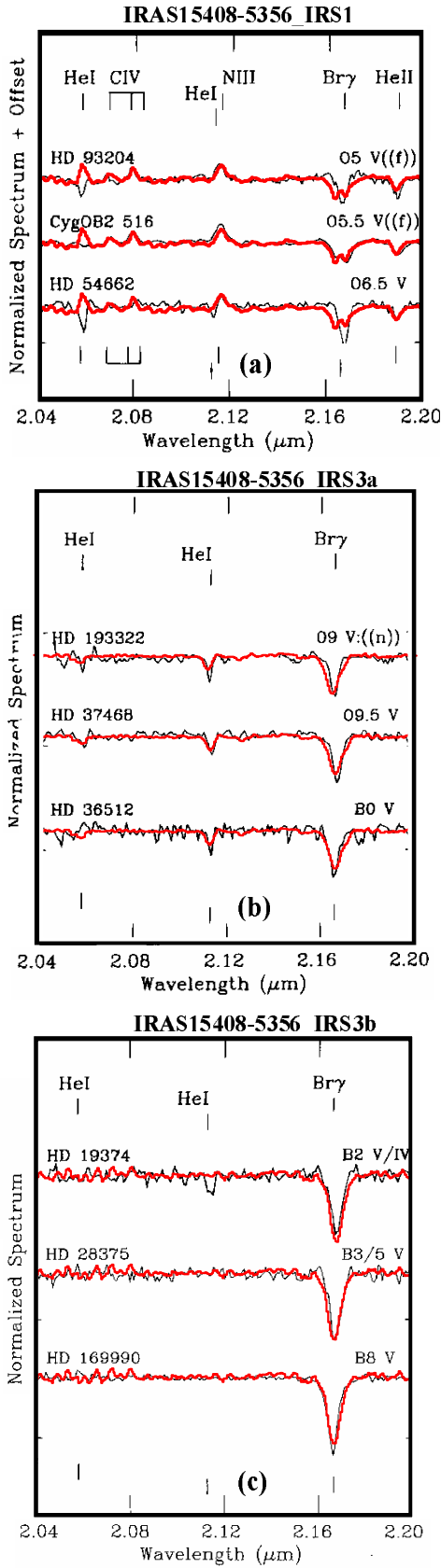}
\caption{GNIRS spectra of IRAS15408-5356 IRS1, IRS3a and IRS3b sources (red lines) overlayed on
$K$ band spectra (black lines) taken from the library of Hanson et al. (1996).}
\end{figure}

\subsection{The IRAS15408-5356 cluster}

In this cluster, spectra of four program stars (IRS1, IRS2, IRS3, and
IRS4) were obtained. The spectrum of IRS1 is shown in Figure 4.
It presents features typical of very hot stars, such as C{\sc iv} (at 2.078 $\mu$m) and N{\sc iii} (at 2.116 $\mu$m) in
emission, characteristic of stars with spectral type earlier than
O6{\sc v}. A direct comparison with the library
templates of Hanson et al. (1996) allowed us to classify it as
O5.5{\sc v} (Figure 8a). Morisset (2004) found that the
relative intensities of mid-infrared emission lines observed by the
{\it Infrared Space Observatory (ISO)}, require an ionizing 
source with effective temperature T$_{\rm eff}=48,700$ K, equivalent
to an O3{\sc v} star, implying that other stars must be contributing
 to the total ionizing luminosity.

IRS3, similar to IRS1, was previously suggested to be one of the main ionising sources of RCW 95 (Roman-Lopes \& Abraham
2004b). 
It actually consists of two stars, labeled IRS3a and IRS3b, as shown in the inset in Figure 1b.
The former presents an unexpected spectrum; it does show the hydrogen Br$\gamma$ and the HeI lines
typical of hot stars, but also the CO band-head overtone lines in absorption,
characteristic of late-type stellar atmospheres. Although CO overtone bands in absorption
have been widely reported in low-mass YSO spectra (Straw, Hyland \& Mc Gregor 1987, Straw et al. 1987, Carr
1989, Casali \& Matthews 1992),  they have also been found  in a few late-O/early-B stars (Hoffmeister et al. 2006). 
The origin of these CO features is not clear; they might be the signature of a cold star in the same line of sight 
or simply neutral gas in the interface between the HII region and the molecular cloud. Apart from the CO lines 
beyond 2.3 $\mu$m,  this star can be classified as 
O9.5{\sc v}, as shown in Figure 8b.  IRS3b has been classified as an B3/B5{\sc v} star (Figure 8c).

Bik et al. (2005) obtained $K$-band spectra for two stars in common with the studied region: 15408nr1410 (O5{\sc v}-O6.5{\sc v})
and 15408nr1454 (O8{\sc v}-B2.5{\sc v}) and, according to their coordinates (Bik 2004), these objects correspond to sources IRS1 (O5.5{\sc v})
and IRS3a (O9.5{\sc v}). Therefore, the spectral classification of the two works agree with each other within the error bars.

The spectrum of IRS2, seen in Figure 6, does not present any evident photospheric features, except for the Br$\gamma$ emission line, superposed 
on an absorption profile. Another emission line near 2.27 $\mu$m might be due to the 4-3 S(4) H$_2$ transition. Differently from 
IRAS 09149-4743 IRS1b,
the spectrum of IRS2 does not show any evidence of a CO environment that could be associated with circumstellar disks. However, since 
this object shows 
large IR color excess and is associated with the bright source MSX6C-G326.6570+00.5912, our result confirms
its previous classification as an YSO (Roman-Lopes \& Abraham 2004b).

Three additional less luminous stars located near IRS2 fell into the slit: IRS10 and other two sources not 
included in the previous photometric study 
of the region, labeled IRS2\_ap2 and IRS2\_ap3.
Their spectra, shown in Figure 5, are characteristic of late-type stars. This result implies that IRS10 is 
not one of the ionising sources of RCW95, 
as previously proposed considering its $JHK$ magnitudes and  colors (Roman-Lopes \& Abraham 2004b). Its location in the colour-colour diagram seems to 
result from its low effective temperature and the high extinction of the cloud, which together mimic the colour indices of young, massive stars.

The spectrum of IRS4, shown in Figure 5, is clearly late-type, with strong CO absorption lines in
the spectral region between 2.29 $\mu$m $ < \lambda < 2.4 \mu$m. One can see  also
a few metallic lines, probably due to NaI and CaI  transitions, which reinforces this classification.

The distances to the early-type sources have been determined from their spectral types, $J$ and $K$  magnitudes, and 
${\rm E}(J-H)$ and ${\rm E}(H-K)$ colour excesses, as described in 
Section 3.4, and presented in Table 3, with the exception of source IRS3b. Since no photometric data are available for this object, we used 
the visual absorption calculated  for IRS3a 
instead, and the $K$ magnitude obtained from the GNIRS $K$-band acquisition image. The instrumental  GNIRS magnitudes were calibrated using the  
combined $K$ magnitudes of IRS3a and IRS3b, obtained from the work of Roman-Lopes \& Abraham (2004b). The resulting individual $K$ magnitudes are 
9.5 and 12.0 for IRS3a and IRS3b, respectively.

The  derived distances to all stars observed in this cluster are similar, like in  IRAS09149-4743, ranging from $1.32\pm0.20$ kpc for IRS3a to 
$1.34\pm0.16$ kpc for IRS1, giving a mean distance of $1.3 \pm 0.2$ kpc for the IRAS15408-5356 cluster.
Giveon et al. (2002) derived a kinematic distance of 2.4 kpc using the Galactic rotation curve, assuming galactocentric distance of 8.5 kpc 
and a solar rotation  velocity of 220 km s$^{-1}$. This is about twice what we found, showing that at least in this direction, the
Galactic kinematic model fails. In fact,
differences between the kinematic distances, derived from radio observations and those obtained by spectroscopic paralaxes were also 
found by Blum et al. (1999, 2000) and Figuer\^edo et al. (2005), showing the importance of spectrophotometric studies for a better understanding 
of the rotation curve of the Galaxy.

\subsection{The clusters in the RCW106 region}

The two clusters associated to IRAS16132-5039 and IRAS16177-5018 are part of the RCW106 complex. One 
program target was observed in each star formation region: IRS3 in the former cluster and IRS1 in the 
latter.

\begin{figure}
\centering
\includegraphics[bb=40 139 299 771,width=6.5cm,clip]{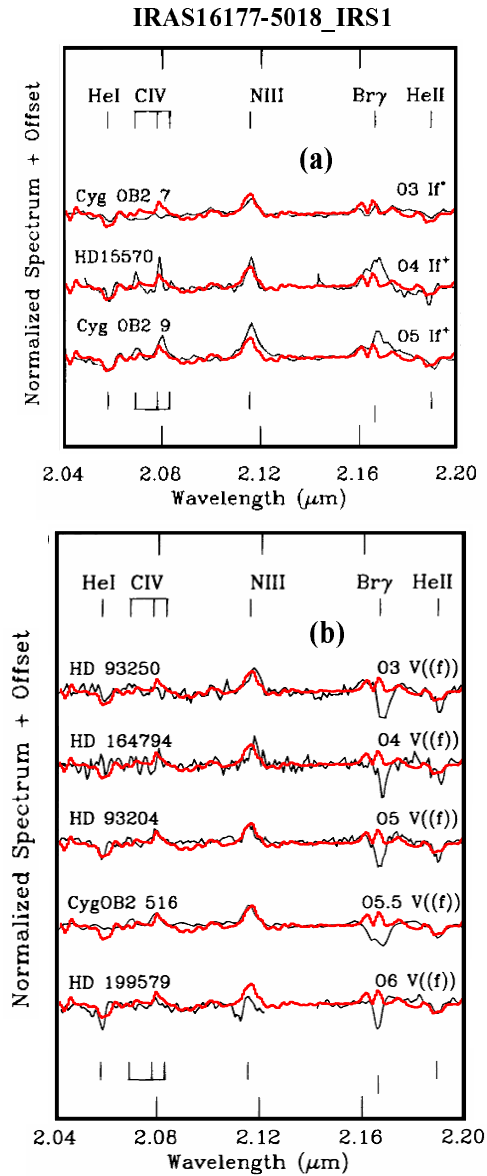}
\caption{GNIRS spectrum of the IRAS16177-5018-IRS1 source (red lines). The spectrum can be compared with those 
of O3-O5{\sc i}f$^+$, and O3-O6{\sc v} stars
taken from the library of Hanson et al. 1996 (black).
All spectra were flux normalized and are set at same scale.}
\end{figure}

The spectrum of the IRS1 source (Fig. 4), shows C{\sc iv}, N{\sc iii}, and N{\sc v} emission lines at 2.078 $\mu$m, 
2.116 $\mu$m, and 2.100 $\mu$m,
respectively, HeII at 2.189 $\mu$m in absorption, and a weak Br$\gamma$ line.
The presence of the N{\sc v} line indicate that IRS1 might be a very hot star. In fact, such line is only found 
in the $K$-band spectra
of O3{\sc v} and O3-O4{\sc i}f$^+$ supergiant stars (Hanson et al. 2005).

In Figure 9a we compare the IRS1 spectrum with that obtained for O3-O5{\sc i}f$^+$ supergiant stars 
(Hanson et al. 1996).
One can notice that the GNIRS spectrum resembles well that of Cyg OB2 7, an O3{\sc i}f$^+$ star.
For completeness, in Fig. 9b we also compare the IRS1 spectrum with that of O3-O6 main-sequence stars.  
In this case we see a reasonable agreement between the GNIRS spectrum features with that from the 
HD93250 (O3{\sc v}), HD164794 (O4{\sc v}), and
HD93204 (O5{\sc v}). The exception are the Br$\gamma$ and
He {\sc ii} lines, which in the templates appear stronger in absorption. 

There are additional constraints indicating that the main ionizing source 
of the compact H{\sc ii} region probably is an extremely hot source. From the K-band spectrum of the associated 
nebular emission, which is shown in Fig. 10,
we can see the presence of strong emission lines like the Br$\gamma$ (2.166 $\mu$m), HeI 
(2.058 $\mu$m, 2.113 $\mu$m and 2.161-2.162 $\mu$m), and [Fe{\sc iii}], which are normally found in HII regions hosting 
embedded hot stars. 
We also found the s-process [Kr{\sc iii}] and [Se{\sc iv}] high excitation emission lines, previously
identified only in planetary nebula (Sterling et al. 2007).
Blum \& McGregor (2008) also detected such lines in their study of the 
ionising stars associated to the UC H{\sc ii} region G45.45+0.06, suggesting that high density H{\sc ii} regions
excited by the hottest O stars also would produce such emission lines.
Indeed, the measured He{\sc i} 2.113/Br$\gamma$ ratio (see Fig. 10) of 0.045$\pm$0.003 (not corrected for reddening) 
indicate the presence of an exciting star whose $T_{eff}$ is greater than 40,000 K (Hanson, Luhman \& Rieke 2002). 
This would correspond to 
a star of spectral type \emph{earlier} than O5{\sc v}, O4.5{\sc iii}, and O4{\sc i} (taking into account the luminosity class), 
as inferred from the new calibration of O star parameters published by Martins et al. (2005).

\begin{figure}
\centering
\includegraphics[bb=56 389 330 720,width=8.5cm,clip]{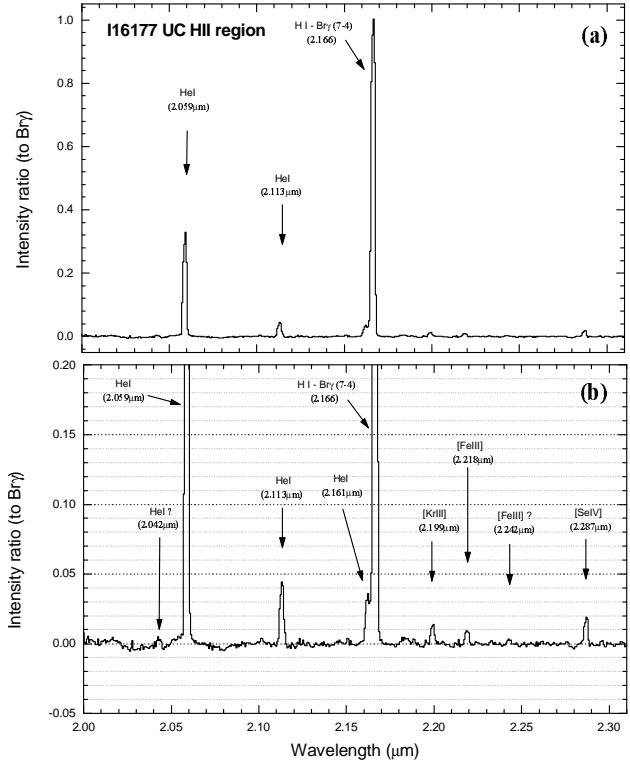}
\caption{(a) GNIRS $K$ band spectra of the I16177 UC HII region. The intensity of the lines were normalized to that
of the Br$\gamma$ one. (b) A detailed view of the spectrum where we can see the presence of the [Fe{\sc iii}]
and the s-process [Kr{\sc iii}] and [Se{\sc iv}] high excitation emission lines.}
\end{figure}

The fact that IRS1 has a K-band spectrum similar to that of Cyg OB2 7, the detection of the s-process 
[Kr{\sc iii}] and [Se{\sc iv}] high excitation emission lines (from the associated UC H{\sc ii} region), 
and the lower limit to the effective
temperature (40,000 K) for the main ionising source in the region, is reasonable to suppose that IRS1 
may be a O3 supergiant star (or alternatively a very hot ionising source). 
Taking into account the rarity of such type of objects, this is an extraordinary
result, which could indicate that RCW106 may be the birthplace of extremely massive stars.
 
The presence of a supergiant star in a young massive stellar cluster could be questioned in view of the 
youth of the sources in the region (Roman-Lopes, Abraham \& Lepine 2003). In fact, 
Roman-Lopes (2007) using the fraction of NIR sources showing excess emission in the NIR, estimated the age of the 
cluster as 2.5-3.0 Myrs.
Indeed, other O-type giants and supergiants have been found in very young stellar clusters.
For instance, Massey et al. (2001) derived ages for several massive early-type stars, founding values in the range 
$1.0-3.0 \times 10^6$ years. Recently, Melena et al. (2008)
detected several early O giant and supergiant stars with similar ages.

Assuming that IRS1 is a O3 {\sc i}f$^+$ supergiant star, we computed its distance as 2.6$\pm$0.7 kpc, compatible with
that of 2.8$\pm$0.6 kpc obtained from the  galactic rotation curve assuming
$R_0=8.5$ kpc, ${\Theta}_{\odot}=220$ km s$^{-1}$ (Honma \& Sofue 1997), and a radial velocity 
$V_R$=-49.5 km s$^{-1}$. On the other hand, considering IRS1 as a O3-O5 {\sc v} star, the distance 
drops to about 1.2$\pm$0.7 kpc. 
Indeed, further $K$-band spectroscopic observations of other nearby NIR sources will be necessary to 
improve our understanding of the Galactic structure in this direction.

\begin{figure}
\centering
\includegraphics[bb=62 237 270 755,width=7cm,clip]{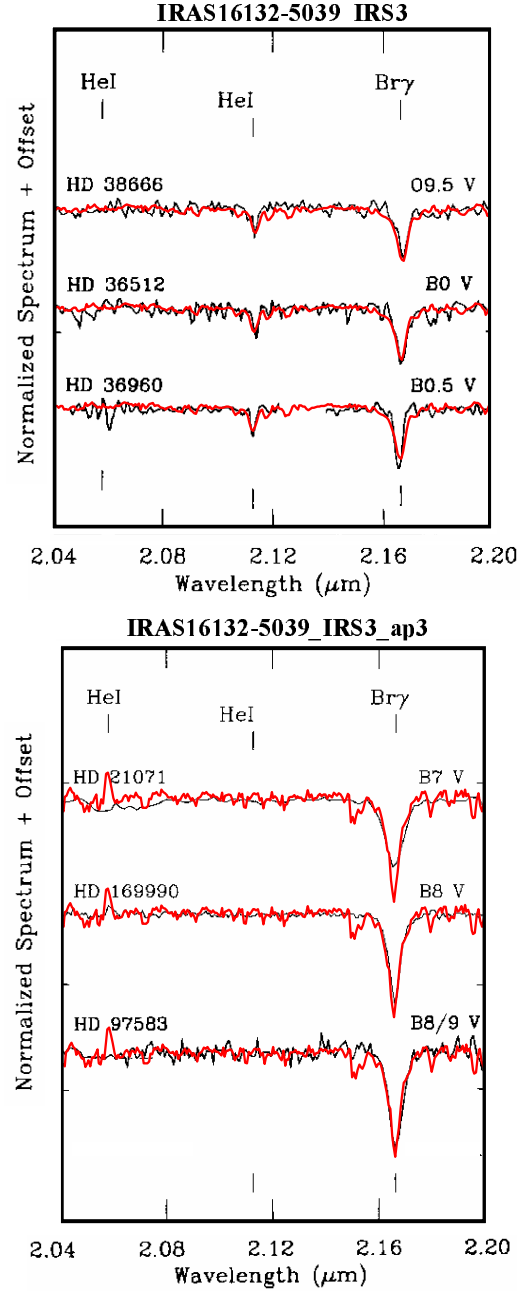}
\caption{GNIRS $K$ band spectra (red lines) of two sources in the cluster associated
to the IRAS16132-5039 source, overlayed on
$K$ band spectra (black lines) taken from the library of Hanson et al. (1996).}
\end{figure}

Bik et al. (2005) also used NIR spectroscopy to estimate the distance to this cluster. Three sources were observed (16177nr271, 16177nr405,
and 16177nr1020), but unfortunately there are no coordinates or other information that would allow us to perform a cross-correlation
between those data with ours. The individual determinations of the distance in that work show somewhat large error bars (1.0 $-$ 3.7 kpc),
but considering its mean value of about 2.4 kpc, their results can be considered consistent with ours.

Another observed point source in this cluster is IRS7 which spectrum does not show any photospheric spectral 
features but the Br$\gamma$ and HeI (2.059 $\mu$m) lines in emission, characteristics of an YSO 
(Figure 6). As pointed out by Roman-Lopes et al. (2003) this object coincides with the mid-infrared source 
MSX6C-G333.3072-00.3666, which reinforces 
this classification.
 We notice a P-Cygni profile in the He{\sc i} line at 2.058 $\mu$m that may indicate the
presence of expanding material inside the region. High-resolution radio observations would be useful
to clarify this issue.

The other studied cluster in RCW106 is IRAS16132-5039.
During the observation of IRAS16132-5039\_IRS3 three stars fell into the slit: IRS3 itself, which has a $K$ band 
spectrum similar to  
O9.5{\sc v}/B0{\sc v} (Figure 10a), and two additional sources: IRS3\_ap2, identified as an YSO  (Figure 5), and IRS3\_ap3, 
probably a B8V/B9V star (Figure 10b).
Similarly to  IRAS15408-5356\_IRS3a, the spectrum of IRAS16132-5039\_IRS3 presents H and He lines as well as CO 
overtone lines; the 
possible origin of the CO lines was already discussed in a previous section.
The distances to IRS3 and IRS3\_ap3 were found to be $2.15\pm 0.50$ and $2.64\pm 0.30$ kpc, respectively (Table 3). 
Therefore, it can be concluded
that both sources are located at the same distance.
The derived spectrophotometric mean distance is about 2.4$\pm$0.5 kpc, a value that is compatible with that 
obtained from the Galactic rotation
curve (3.2$\pm 0.6$ kpc) using the radial velocities taken from the CS (2-1) line (Bronfmann et al. 1996) and from the 
hydrogen recombination lines (Caswell \& Haynes, 1987).

\section{Summary}

We report  $K$-band spectroscopic observations of stars in highly reddened  young 
stellar clusters, obtained with GNIRS. We found 8 massive main-sequence stars, 1 possible supergiant, 
7 late-type stars and 4 YSOs.

The main-ionising star in the cluster associated to IRAS09149-4743,  is IRS2 (O9{\sc v}), which together with 
IRS1 (B0{\sc v}) are the dominant ionising sources in the RCW41 HII region. They are located in cloud A of the VMR at 
a mean distance of 1.3$\pm$0.2 kpc, in agreement with 
the value derived from the kinematic method using CO radial velocities.
We found that IRS09149\_IRS1   actually  consists of two sources: a  B0{\sc v} star (IRS1a) and  an YSO (IRS1b). This 
result allows us to state that the 
infrared excess found by Ortiz et al. (2007), comes from the dusty nearby companion of IRS1a. The other bright 
source in the region,
IRS09149\_IRS3, is a background giant star located at least  twice as far as the cluster, confirming the assumption 
made
by Ortiz et al. (2007), who suggested that this source is a late-type star.

Four $K$-band spectra were obtained in the cluster associated with RCW95. IRAS15408\_IRS1  has a $K$-band
spectrum compatible with an O5.5{\sc v} star, while IRS3a and IRS3b were classified as O9.5{\sc v} and B3-B5{\sc v}, respectively. 
Their distances are similar and have an average value of 1.3$\pm$0.2 kpc,  
 in disagreement with the kinematic distance of  2.4 kpc. This seems to indicate that, at least for this
star forming region, the
galactic kinematic model fails. Eventually, these results can be used to improve our understanding of the rotation 
curve of the Galaxy.
IRS2, another very bright source in this cluster,
is associated with the mid-IR source MSX6C-G326.6570+0.5912, and was classified by Roman-Lopes \& Abraham (2004b) 
as an YSO candidate.
Its $K$-band spectrum presents Br$\gamma$ and HeI emission lines characteristic of YSO´s, confirming the previous 
classification.

One of the sources observed in the direction of RCW106  belongs to the cluster associated with 
IRAS16132-5039. Three other stars fell into the spectrograph slit: IRAS16132\_IRS3 (O9.5-B0{\sc v}), IRAS16132\_IRS3\_ap2 
(YSO) 
and IRAS16132\_IRS3\_ap3 (B8-B9{\sc v}).
The derived mean distance is 2.4$\pm$0.5 kpc, a value compatible with that obtained from the kinematic method (3.2$\pm 
0.6$ kpc)
using the CS (2-1) and hydrogen recombination lines. 

The other cluster in the RCW106 region is associated with IRAS16177-5018. The IRAS16177\_IRS1 spectrum presents 
C{\sc iv}, N{\sc iii}, and N{\sc v} emission lines at 2.078 $\mu$m, 2.116 $\mu$m, and 2.100 $\mu$m respectively, 
which are seen only in very hot stars. 
We also detected in the associated nebular K-band spectrum, the s-process [Kr{\sc iii}] and [Se{\sc iv}] 
high excitation emission lines, previously identified only in planetary nebula. In the case of H{\sc ii} regions, 
such lines seem to be produced only in high density environment excited by the hottest O stars.
Indeed, the measured He{\sc i} 2.113/Br$\gamma$ ratio 
indicate the presence of an exciting star whose $T_{eff}$ is greater than 40,000 K. This would correspond to 
a single star of spectral type
\emph{earlier} than O5{\sc v}, or O4.5{\sc iii}, or O4{\sc i}, depending on the luminosity class.

We found a good agreement between the IRAS16177\_IRS1 $K$-band spectrum and that 
of Cyg OB2 7, an O3If* star. Taking into account the scarcity of supergiants in young cluster (though a 
few similar occurrences have been reported elsewhere), this is an 
extraordinary result, which could indicate that RCW106 may be the birthplace of extremely massive stars. 

Considering IRAS16177\_IRS1 as an O3If* star, the distance to the cluster would be 2.6$\pm$0.7 kpc,  
similar to that of the other massive star formation region in the RCW106 complex.
On the other hand, if IRAS16177\_IRS1 is an O3-O5 main-sequence star, its distance drops 
to  1.2$\pm$0.7 kpc, much smaller than its kinematic distance.

We originally planned to obtain the $K$-band spectrum for one  target in the IRAS16177-5018 region, but another 
source also fell into the slit: IRAS16177\_IRS7. Its spectrum  is typical of an YSO, in agreement with the  
classification proposed by Roman-Lopes, Abraham \& Lepine (2003). Its HeI 
(2.058 $\mu$m) emission line shows a P-Cygni profile that may indicate the presence of expanding 
motion. High resolution radio observations would be useful to clarify this issue.

\section*{Acknowledgments}

We thank the referee, Willem-Jan de Wit, for
his suggestions and comments which contributed to improve
the presentation of the paper.
This work was partially supported by the Brazilian agencies CNPq and FAPESP. A. 
Roman-Lopes thanks financial support from FAPESP under the program 04/10375-2, and CONICYT (Chile)
program 31060004. A.R.-A. 
acknowledges support from the Brazilian Agency CNPq under program 
311476/2006-6. Based on observations obtained at the Gemini Observatory, which is 
operated by the Association of Universities for Research in Astronomy, 
Inc., under a cooperative agreement with the NSF on behalf of the Gemini 
partnership: the National Science Foundation (United States), the Science 
and Technology Facilities Council (United Kingdom), the National Research 
Council (Canada), CONICYT (Chile), the Australian Research Council 
(Australia), Minist\'erio da Ci\^encia e Tecnologia (Brazil) and SECYT 
(Argentina).

\begin{table*}
 \centering
 \begin{minipage}{160mm}
  \caption{Log-book of the GNIRS observations. The columns are: (1) The associated IRAS source; (2) the identifier of the point
  source, as 
  designated in the original photometric survey;(3,4) equatorial coordinates (J2000.0);(5) the total exposure time (s);(6)the mean airmass 
  at the time of the observations;(7) the Hipparcos identifier of the associated A0V telluric star;(8) the mean airmass of the telluric star 
  at the time of the observations;(9) lists the Gemini identification program and Column (10) the corresponding date of observation.}
  \begin{tabular}{@{}cccccccccc@{}}
  \hline

Source & IRS & $\alpha$ (J2000.0)& $\delta$ (J2000.0) & Itime (s) & X & Telluric & X & ID Program & Date\\
(1) & (2) & (3) & (4) & (5) & (6) & (7) & (8) & (9) & (10)\\

09149-4743 & 1  & 09h16m43.50s & -47$^o$56'23.0''  & 540 & 1.07 & HIP40974 & 1.03 & GS-2005B-Q-33 & Dec 23, 2005\\
09149-4743 & 2  & 09h16m47.94s & -47$^o$57'18.0''  & 132 & 1.06 & HIP40974 & 1.05 & GS-2005B-Q-33 & Dec 23, 2005\\
09149-4743 & 3  & 09h16m41.89s & -47$^o$56'15.4''  & 540 & 1.05 & HIP40974 & 1.06 & GS-2005B-Q-33 & Dec 24, 2005\\
15408-5356 & 1  & 15h44m43.40s & -54$^o$05'53.7''  & 180 & 1.30 & HIP75161 & 1.36 & GS-2005B-Q-33 & Dec 24, 2005\\
15408-5356 & 2  & 15h44m43.40s & -54$^o$05'53.7''  & 180 & 1.34 & HIP75161 & 1.35 & GS-2005B-Q-33 & Dec 27, 2005\\
15408-5356 & 3  & 15h44m56.17s & -54$^o$07'18.1''  & 360 & 1.57 & HIP75161 & 1.63 & GS-2005B-Q-33 & Jan 22, 2006\\
15408-5356 & 4  & 15h44m53.04s & -54$^o$06'31.9''  & 1080 & 1.64 & HIP75161 & 1.63 & GS-2005B-Q-33 & Jan 23, 2006\\
16132-5039 & 3 & 16h16m55.99s & -50$^o$47'22.8''  & 1080 & 1.88 & HIP83406 & 1.75 & GS-2005B-Q-43 & Sep 25, 2004\\
16177-5018 & 1 & 16h21m31.60s & -50$^o$25'08.3''  & 1080 & 1.56 & HIP83818 & 1.65 & GS-2005B-Q-33 & Jan 24, 2006\\
\hline
\end{tabular}
\end{minipage}
\end{table*}

\begin{table*}
\centering
\begin{minipage}{180mm}
\caption{Derived spectral types for the hot stars in our sample.}
\begin{tabular}{ccccccccccc}
\hline


Source & $\alpha$ (J2000)& $\delta$ (J2000)& S.T. & M$_{J}$ & (J-H)$_o$ & (H-K)$_o$ & J & (J-H) & K & (H-K) \\
(1) & (2) & (3) & (4) & (5) & (6) & (7) & (8) & (9) & (10) & (11) \\

\hline

I09149\_IRS1a & 09h16m43.50s & -47$^o$56'23.0'' & B0V & -2.21 & -0.16 & -0.01 & 10.70 & 0.80 & 9.40 & 0.50 \\
I09149\_IRS2 & 09h16m47.94s & -47$^o$57'18.0'' & O9V & -3.00 & -0.18 & -0.01 & 9.72 & 0.66  & 8.68 & 0.38 \\
I09149\_IRS31 & 09h16m47.77s & -47$^o$57'15.1'' & B7/B8V & -0.27/0.29 & -0.07 & +0.02 & 13.34 & 0.80 & 12.25 & 0.29 \\

I15408\_IRS1 & 15h44m43.40s & -54$^o$05'53.7'' & O5.5V & -4.05 & -0.20 & -0.01 & 10.76 & 1.37 & 8.57 & 0.82 \\
I15408\_IRS3a & 15h44m56.17s & -54$^o$07'18.1'' & O9.5V & -2.85 & -0.16 & -0.01 & 11.25 & 1.17  & 9.50 & 0.58 \\
I15408\_IRS3b$^1$ & 15h44m56.17s & -54$^o$07'18.1'' & B3/B5V & -0.45/+0.15 & -0.11 & +0.01 & --- & ---  & 12.0 & --- \\

I16132\_IRS3 & 16h16m56.01s & -50$^o$47'22.7'' & O9.5/B0V & -2.85/-2.21 & -0.18/-0.16 & -0.01 & 11.77 & 0.76  & 10.30 & 0.71 \\
I16132\_IRS3\_ap3 & 16h16m50.84s & -50$^o$47'44.2'' & B8/B9V & -0.27/0.29 & -0.03 & -0.01 & 12.71 & 0.15 & 12.50 & 0.06 \\

I16177\_IRS1 & 16h21m31.60s & -50$^o$25'08.3'' & O3If*/O3-O5V & -6.85/-4.95 & -0.21 & -0.01 & 16.65 & 3.92 & 10.21 & 2.52\\

\hline
\hline
\end{tabular}
\end{minipage}
\end{table*}

\begin{table*}
\centering
\begin{minipage}{180mm}
\caption{Summary of the extinction values and distances to the program stars. For the I16177\_IRS1 source we show the distances obtained considering
class I (1) and class V (2) case.}
\begin{tabular}{lcccccccccccc}
\hline 

Source & $E(J-H$) & $A_J$  & $A_J$  & $A_J$  & $A_V$ & $E(H-K$) & $A_K$  & $A_K$  & $A_K$  & $A_V$ & $d$  & $d_{\rm cluster}^2$  \\
& & $R=2.8$ & $R=3.1$ & $R=5.0$ & $R=3.1 $ & & $R=2.8$& $R=3.1$& $R=5.0$& $R=3.1$&(kpc) & (kpc)\\
(1) & (2) & (3) & (4) & (5) & (6) & (7) & (8) & (9) & (10) & (11) & (12) & (13) \\
\hline

I09149\_IRS1a & 0.96 & 2.5 & 2.6 & 2.8 & 9.1 & 0.51 & 1.0 & 1.0 & 1.1 & 8.9 & $1.20 \pm{0.12}$ & $1.3 \pm{0.2}$ \\
I09149\_IRS2 & 0.84 & 2.2 & 2.3 & 2.5 & 8.0 & 0.39 & 0.9 & 0.9 & 1.0 & 6.8 &$1.27\pm{0.13}$ \\
I09149\_IRS31 & 0.87 & 2.3 & 2.3 & 2.6 & 8.3 & 0.27 & 0.9 & 0.9 & 1.0 & 4.7 &$1.37\pm{0.19}$ &  \\
I15408\_IRS1 & 1.57 & 4.1 & 4.2 & 4.6 & 14.9 & 0.83 & 1.6 & 1.7 & 1.9 & 14.4 &$1.34\pm{0.16}$& $1.3 \pm{0.2}$ \\
I15408\_IRS3a & 1.33 & 3.5 & 3.6 & 3.9 & 12.6 & 0.59 & 1.4 & 1.4 & 1.6 & 10.2 &$1.32\pm{0.20}$ \\
I15408\_IRS3b & * & * & * &*&  12.6 & * & & * & * & 10.2 &$1.24\pm{0.21}$ \\
I16132\_IRS3 & 0.92 & 2.4 & 2.5 & 2.7 & 8.7 & 0.72 & 1.0 & 1.0 & 1.1 & 12.5 &$2.15\pm{0.50}$ & $2.4 \pm{0.5}$ \\
I16132\_3(ap3) & 0.22 & 0.6 & 0.6 & 0.7 & 2.1 & 0.04 & 0.2 & 0.2 & 0.3 & 0.7 &$2.64\pm{0.30}$ &  \\
I16177\_IRS1$^{(1)}$ & 4.13 & 10.7 & 11.1 & 12.1 & 39.2 & 2.53 & 4.3 & 4.5 & 4.9 & 43.9 &$2.58\pm{0.68}$ & $2.6 \pm{0.7}$ \\
I16177\_IRS1$^{(2)}$ & 4.13 & 10.7 & 11.1 & 12.1 & 39.2 & 2.53 & 4.3 & 4.5 & 4.9 & 43.9 &$1.22\pm{0.74}$ & $1.2 \pm{0.7}$ \\

\hline
\hline
\end{tabular}
\end{minipage}
\end{table*}

\end{document}